\documentstyle[twocolumn,epsf]{jpsj}
\newcommand{\navo}{\mbox{$\rm\alpha'$-$\rm NaV_2O_5$}}
\newcommand{\Tc}{\mbox{$T_{\rm C}$}}
\newcommand{\SG}[1]{\mbox{$\it #1$}}

\def\runtitle{LETTERS}
\def\runauthor{Tetsuo {\sc Ohama} {\it et al.}}

\title{Zigzag Charge Ordering in $\alpha'$-NaV$_2$O$_5$}

\author{Tetsuo {\sc Ohama}$^{1,}$\footnote
{E-mail: ohama@physics.s.chiba-u.ac.jp},
Atsushi {\sc Goto}$^2$, Tadashi {\sc Shimizu}$^2$, Emi {\sc Ninomiya}$^3$,
Hiroshi {\sc Sawa}$^3$, Masahiko {\sc Isobe}$^4$ and Yutaka {\sc Ueda}$^4$} 
\inst{$^1$Department of Physics, Faculty of Science, Chiba University, Chiba
263-8522\\ $^2$National Research Institute for Metals, Tsukuba 305-0003\\
$^3$Graduate School of Science and Technology, Chiba University,
Chiba 263-8522\\
$^4$Institute for Solid State Physics, University of Tokyo,
Roppongi, Minato-ku, Tokyo 106-8566}

\recdate{}

\abst
{$^{23}$Na NMR spectrum measurements in \navo\ with a single-crystalline
sample are reported.
In the charge-ordered phase, 
the number of inequivalent Na sites observed
is more than that expected from the low-temperature structures
of space group \SG{Fmm2}\ reported so far.
This disagreement indicates that the real structure including both
atomic displacement and charge disproportionation is of lower symmetry.
It is suggested that zigzag ordering is the most probable.
The temperature variation of the NMR spectra near the transition temperature
is incompatible with that of second-order transitions.
It is thus concluded that the charge ordering transition is first-order.}

\kword{$\alpha'$-NaV$_{\sf 2}$O$_{\sf 5}$, NMR, charge ordering,
crystal structure, spin-ladder, trellis lattice}

\begin{document}\sloppy\maketitle
Since the phase transition into a spin-gapped phase
in \navo\ was reported,~\cite{Isobe96}
a lot of experimental efforts have been devoted to understand
the nature of this transition.
Although it was initially identified as a spin-Peierls transition,
a recent room-temperature structural study~\cite{Meetsma} questioned this
interpretation:
it concluded that all V ions are in a uniform oxidation state of V$^{4.5+}$
and form a quarter-filled trellis lattice composed of 
two-leg ladders.
After that,
$^{51}$V NMR measurements~\cite{Ohama99} revealed 
charge ordering of V$^{4+}$ and V$^{5+}$ states below
the transition temperature $T_{\rm C}\sim$ 34 K.
Subsequent theoretical studies showed that long-range Coulomb interaction
can induce charge ordering in a quarter-filled trellis lattice.\cite{Seo98,
Nishimoto,Thalmeier,Mostovoy}
These studies suggest zigzag or linear chain ordering
depending on the strength of the long-range Coulomb interactions.
The proposed mechanism of charge ordering is similar to that for
charge density wave in quarter-filled systems of
low-dimensional organic compounds,\cite{SeoRev}
suggesting some common physics to \navo\ and these systems.

Soon after the finding of the transition,
an x-ray diffraction measurement revealed superlattice formation of
$2a\times 2b\times 4c$ in the charge-ordered phase,\cite{Fujii} 
but the detailed low-temperature structure has been unknown yet.
Recently, two x-ray diffraction studies of the low-temperature structure 
were reported.
These indicate almost the same structure  of space group \SG{Fmm2},
but their assignments of V electronic states are different.
One suggests a structure consists of half-filled (V$^{4+}$)
and empty (V$^{5+}$) ladders.~\cite{Luedecke}
This charge distribution disagrees with a recent x-ray anomalous
scattering measurement, which indicates charge modulation along
$b$ axis.\cite{Nakao00}
The other suggests a structure including three different electronic state of
V$^{4+}$, V$^{5+}$ and V$^{4.5+}$.~\cite{Boer}
This structure is incompatible with
the $^{51}$V NMR measurement,~\cite{Ohama99} which clearly shows that
all the V sites split into two groups of V$^{4+}$ and V$^{5+}$ states
and that no V sites remains to be V$^{4.5+}$.
Thus, the low-temperature structure and the charge ordering
pattern are still under discussion.

In this letter, we report $^{23}$Na NMR spectrum measurements with
a single-crystalline sample.
The obtained NMR spectra in the charge-ordered phase
disagree with the space group \SG{Fmm2}, indicating lower symmetry.
We will discuss possible low-temperature structures and charge ordering
patterns, and will show that zigzag patterns are the most probable.
The temperature variation of the NMR spectra near \Tc\
is incompatible with that of second-order transitions,
suggesting that the charge ordering is first-order.

The single-crystalline sample preparation
was described in ref.~\citen{Isobe97}.
The NMR measurements were done using a high-resolution NMR spectrometer
with a magnetic field $H\sim$ 63.382 kOe.
The $^{23}$Na NMR spectra were obtained as power spectra 
by Fourier-transforming FID signals
with the Gaussian multiplication for resolution enhancement and
apodization.~\cite{Sanders}
Line width and intensity in the obtained spectra are thus inaccurate.

\begin{figure}[tb]
  \begin{center}
\epsfxsize=80mm \epsfbox{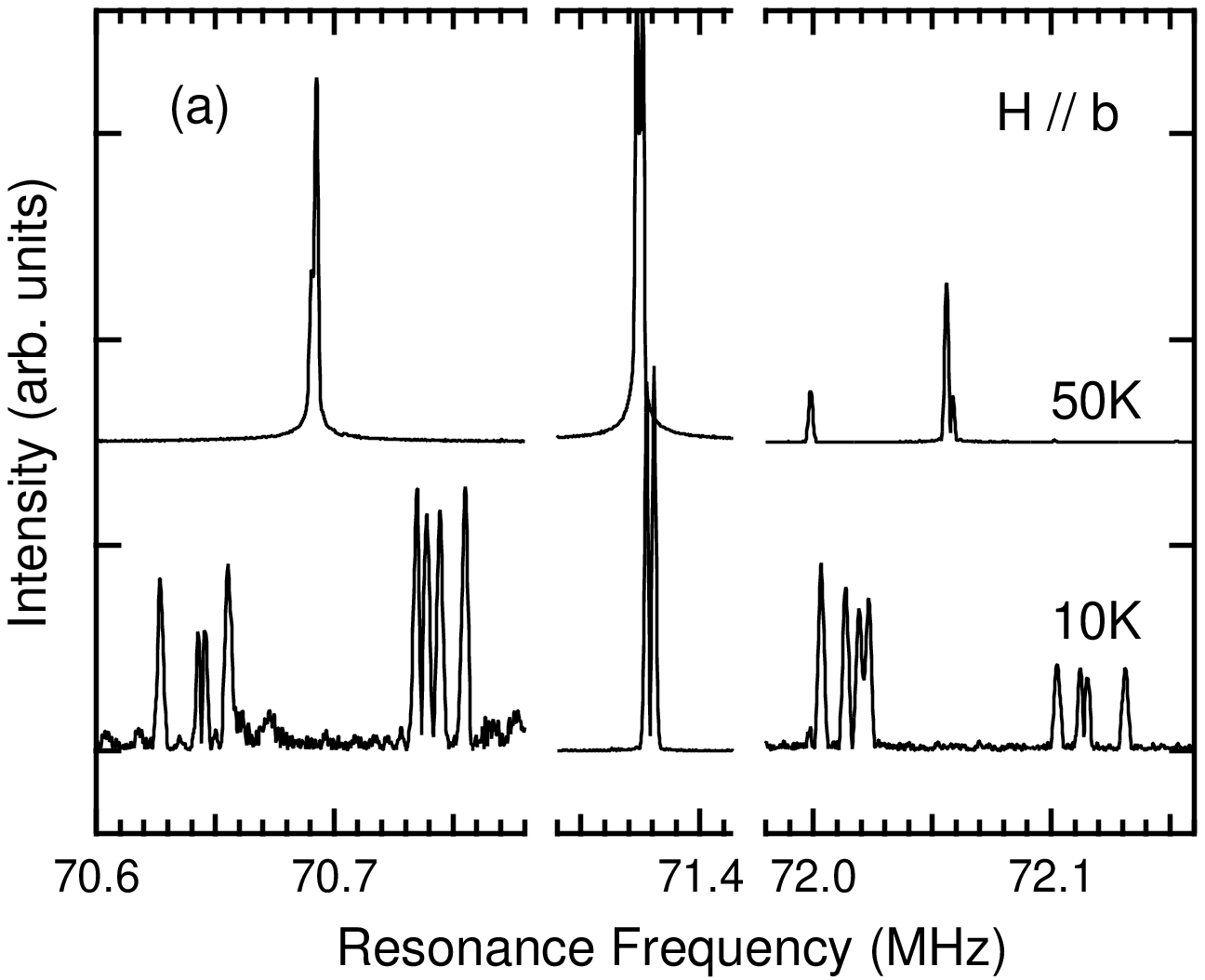}
  \end{center}
  \begin{center}
\epsfxsize=80mm \epsfbox{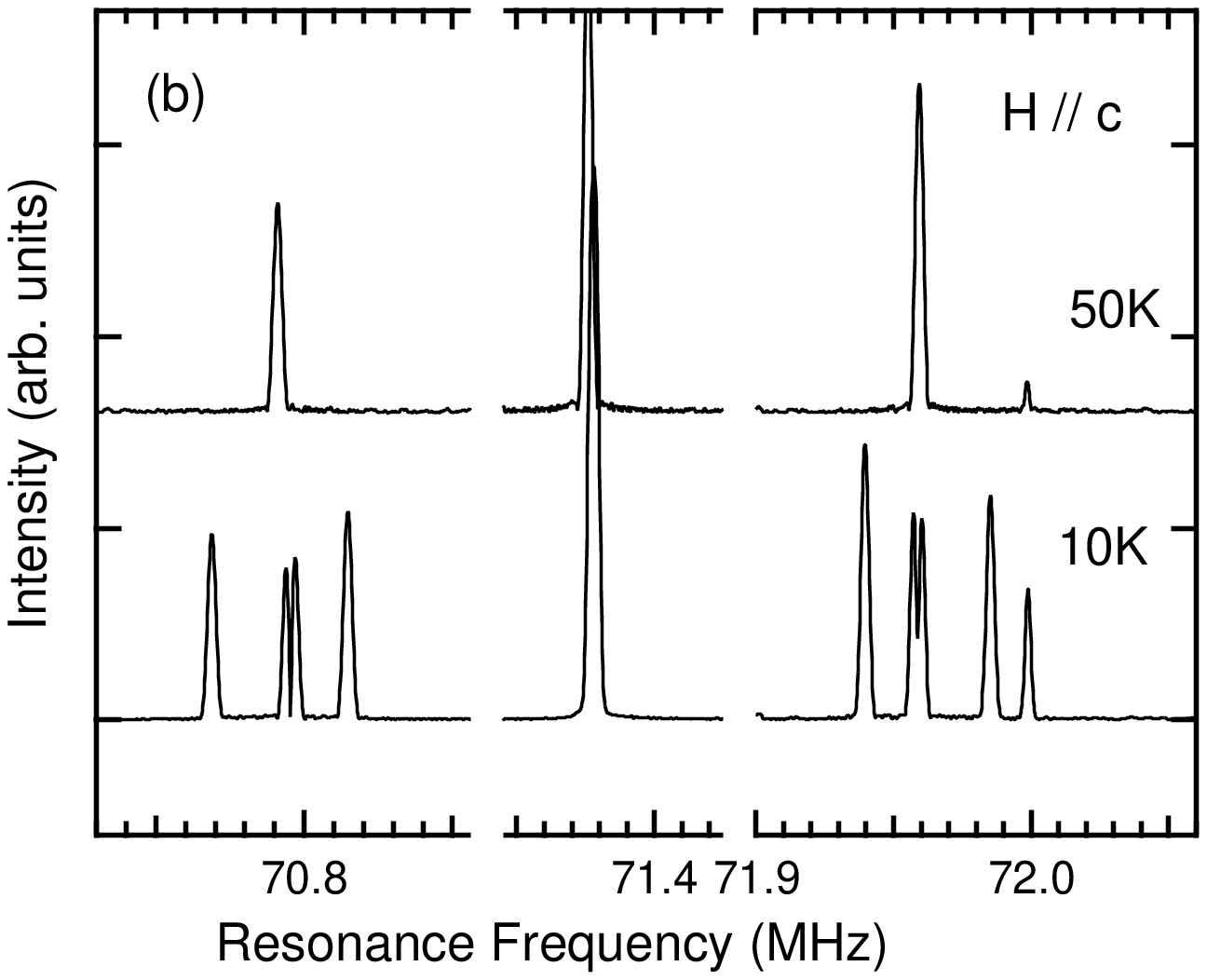}
  \end{center}
\caption{$^{23}$Na NMR spectra at 10 and 50 K for (a) $H\parallel b$ and (b)
$H\parallel c$. Lines observed at 71.9994MHz are an external noise.}
\label{fig:specBC}
\end{figure}
\begin{figure}[tb]
  \begin{center}
\epsfxsize=80mm \epsfbox{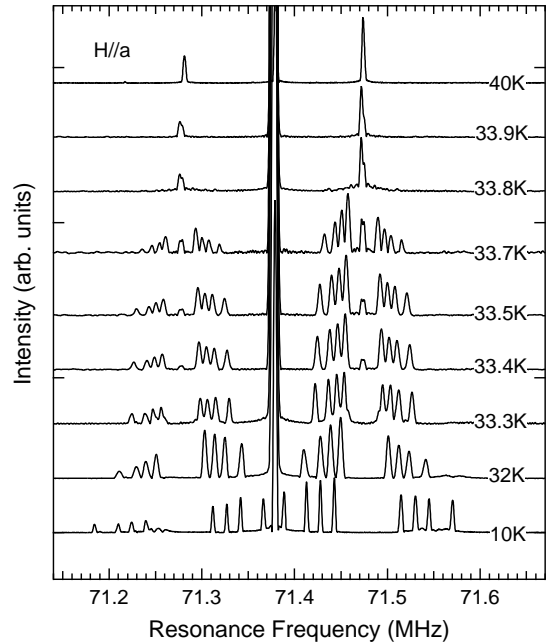}
  \end{center}
\caption{Temperature variation of $^{23}$Na NMR spectra near \Tc\ for 
$H\parallel a$.}
\label{fig:specA}
\end{figure}
NMR spectrum in a crystal is generally determined by 
the number of crystallographically inequivalent nuclear sites,
and electrical field gradient and NMR shift tensors at each site.
These are closely related to the crystal structure and the site symmetry.
The room-temperature structure of \navo\ belongs to space group
$Pmmn$,\cite{Meetsma,Smolinski,Schnering}
in which structure Na ions occupy a unique atomic position
of site symmetry \SG{mm2}.
The principal axes of 
the electrical field gradient and NMR shift tensors are identical to
the crystallographic axes accordingly.
For the low-temperature structure of space group \SG{Fmm2},\cite{Luedecke}
Na ions occupy six different atomic positions.
Among 32 Na atoms in a unit cell,
16 atoms occupy four positions ($4a$) of site symmetry \SG{mm2},
and the other 16 do two positions ($8d$) of site symmetry \SG{.m.}.
For the $4a$ sites, the principal axes are identical to
the crystallographic axes,
whereas for the $8d$ sites,
one of the principal axes is identical to $b$ axis,
but the other two are not determined only by the site symmetry.

We present $^{23}$Na NMR spectra in the charge-ordered and uniform
(above \Tc ) phases for
$H\parallel b$, $c$, and $a$ in Figs.~\ref{fig:specBC}(a),
\ref{fig:specBC}(b), and \ref{fig:specA}, respectively. 
In the spectra in the uniform phase,
three resonance lines (the central and two satellite lines) are observed
corresponding to $^{23}$Na nuclear spins ($I=3/2$) at the unique Na position.
Quadrupolar splitting, nuclear quadrupolar frequency $\nu_{\rm Q}$, and 
quadrupolar asymmetry parameter $\eta$ at 50 and 295 K are deduced 
(Table \ref{table:nuQ_HT}),
in agreement with a previous measurement with
a powder sample.\cite{Ohama97}
\begin{table}[tb]
\caption{Quadrupolar splitting and  $\nu_{\rm Q}$ in kHz and 
$\eta$ in the uniform phase.}
\label{table:nuQ_HT}
\begin{tabular}{rcrrcl}\hline
\multicolumn{1}{c}{$T$(K)} & \multicolumn{1}{c}{$\nu_b\ (\equiv\nu_Q)$} &
\multicolumn{1}{c}{$\nu_a$} & \multicolumn{1}{c}{$\nu_c$}& \multicolumn{1}{c}
{$\nu_b - (\nu_a + \nu_c)$} & \multicolumn{1}{c}{$\eta$}\\ \hline
50 & 682 & 99 & 581 & 2 & 0.85 \\
295 & 641 & 120 & 520 & 1 & 0.62\\
\hline\end{tabular}
\end{table}
Since electrical field gradient is a traceless tensor,
and the principal axes at the Na site
are identical with the crystallographic axes,
$\nu_b - (\nu_a + \nu_c)$ should vanish.
This is confirmed in Table \ref{table:nuQ_HT},
indicating accurate alignments of the crystallographic axes along the
magnetic field.

We observed slight splitting of the resonance lines for $H\parallel b$ above
\Tc .
The splitting of the central line remains unchanged down to 10 K.
We have no plausible explanation for its origin.
For the Na site in the \SG{Pmmn}\ structure,
misalignment of the applied magnetic field
from the crystallographic axes cannot cause 
any line splitting.
Further, since the splitting of the central and satellite lines is
comparable, 
it is not of quadrupolar origin and thus can be ascribed to 
neither inequivalent Na sites nor a twin sample.

In the charge-ordered phase, each satellite line splits into eight lines
for $H\parallel a$ and $b$, or four for $H\parallel c$.
A similar result has been reported in ref.~\citen{Fagot}.
As listed in Table \ref{table:nuQ_LT},
these satellite lines can be assigned to eight Na sites with
different quadrupolar splitting.
\begin{table}[tb]
\caption{Quadrupolar splitting in kHz at 10 K.}
\label{table:nuQ_LT}
\begin{tabular}{ccrrcl}\hline
site & \multicolumn{1}{c}{$\nu_b$} &
\multicolumn{1}{c}{$\nu_a$} & \multicolumn{1}{c}{$\nu_c$}& \multicolumn{1}{c}
{$\nu_b - (\nu_a + \nu_c)$} \\ \hline
A & 752 & 193 & 563 & -4 \\
B & 736 & 153 & 583 & 0 \\
C & 733 & 168 & 563 & 2 \\
D & 724 & 137 & 583 & 4 \\
E & 644 & 65 & 580 & -1 \\
F & 640 & 36 & 608 & -4 \\
G & 634 & 51 & 580 & 3 \\
H & 624 & 11 & 608 & 5 \\
\hline\end{tabular}
\end{table}
Each of the four lines for $H\parallel c$ was considered as superposition
of two lines,
since these four lines have comparable intensities.
This assignment in Table \ref{table:nuQ_LT} is not unique
and other assignments of satellite lines 
for $H\parallel c$ are possible: two adjacent lines with $\nu_c$ of
580 and 583 kHz can be assigned to the sites B, D, E and G alternatively.

The observation of the eight inequivalent Na sites
disagrees with the low-temperature
structures of \SG{Fmm2}, which has the six Na sites.
Crystallographically equivalent sites, in general, can be observed 
as split resonance lines in NMR spectrum for some crystal structures.
However, this is not the case for the Na sites in the \SG{Fmm2} structure
with the magnetic field along the crystallographic axes.
To explain this disagreement between the NMR and x-ray measurements,
it is reasonable to suppose
that the space group determined by the x-ray diffraction\cite{Luedecke}
are correct as far as major atomic displacement is concerned,
and further that the charge disproportionation of V ions is responsible
for the disagreement,
since usual x-ray diffraction measurements are sensitive
to atomic displacement
but not to charge disproportionation of V ions.\cite{Nakao00}
This supposition is supported by the fact that
\SG{Fmm2} contains all the subgroups of \SG{Pmmn},
the room-temperature structure,
with the unit cell of $2a\times 2b\times 4c$. 
Since the observed atomic displacement through the phase transition is small, 
it is unlikely that a new symmetry element arises at the transition.
Then the real low-temperature structure consistent with the present NMR
result should belong to a proper subgroup of \SG{Fmm2}.
\begin{table}[tb]
\caption{Possible subgroups of \SG{Fmm2}\ (order 16)
for low-temperature structure and charge ordering patterns.
A: alternating chains along $a$ axis, C: four-V$^{4+}$ clusters, and
Z1, Z2, Z3: zigzag chains.}
\label{table:SG}
\begin{tabular}{cc}\hline
order & subgroups \\\hline
8 & \SG{Ama2}(C) \SG{Bma2}(C) \SG{Ccc2}(Z1)\\ 
4 & \SG{Pna2_1}(A, C, Z2) \SG{Pbn2_1}(C, Z2, Z3)\\
& \SG{Pca2_1}(A, C, Z2) \SG{Pbc2_1}(C, Z2, Z3)\\  
\hline\end{tabular}
\end{table}

\begin{figure}[tb]
  \begin{center}
\epsfxsize=82mm \epsfbox{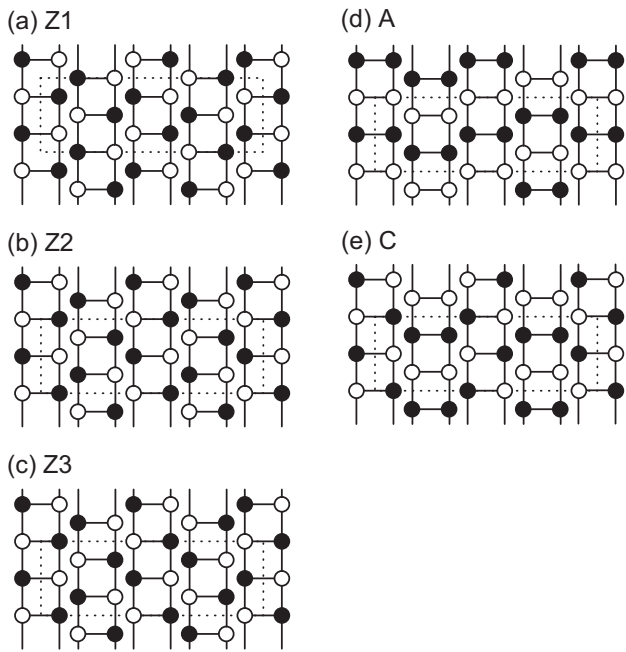}
  \end{center}
\caption{Possible charge ordering patterns
 in a plane of trellis lattice.
(a), (b), (c): possible zigzag patterns (Z1, Z2, Z3),
(d), (e): examples of alternating chain (A) and four-V$^{4+}$ cluster (C)
patterns.
Closed and open circles denote V$^{4+}$ and V$^{5+}$ sites, respectively.
Dashed lines show a unit cell.}
\label{fig:CO}
\end{figure}

In Table \ref{table:SG},
the subgroups compatible with the NMR spectra
and possible charge ordering patterns in a plane of trellis
lattice are listed.
The following subgroups and patterns are here excluded:
(1) the subgroups which are a proper subgroup of the subgroups
of \SG{Fmm2} listed in Table \ref{table:SG}\ and 
include only structures with more than eight Na sites,
since there is no experimental evidence for such lower symmetry,
(2) the patterns without charge modulation along $b$ axis,
which disagree with the x-ray anomalous scattering measurements,\cite{Nakao00}
(3)  the patterns expected to contain V$^{4+}$ sites
with very different magnetic properties, for example, a pattern
containing both zigzag and linear chains, since
all the V$^{4+}$ sites have the same nuclear relaxation rate
well below \Tc .\cite{Ohama99}
The remaining possible patterns are classified in two groups according to
the arrangement of V$^{4+}$ and V$^{5+}$ sites in  two-leg ladders:
(1) the patterns in which a part of rungs of ladders have both the V sites
occupied and consequently another part of rungs are empty.
This group consists of several patterns of alternating chains along $a$
axis (A) and of clusters composed of four V$^{4+}$ sites (C).
(2) the patterns which contain no doubly occupied rung. This group consists of
three types of zigzag patterns (Z1, Z2, Z3).
These zigzag patterns and examples of the patterns with doubly occupied
rungs are shown in Figs.~\ref{fig:CO}(a)--\ref{fig:CO}(e).

The double occupancy of the rungs contained in the former group of charge
ordering patterns
is unfavorable because it has higher energy per rung by
the order of the hopping parameter along the rung $t_\perp\sim$
0.3 eV.\cite{Horsch,Nishimoto}
We therefore concluded that the zigzag patterns are the most probable.
Although this conclusion is consistent with neutron scattering\cite{Yosi}
and dielectric susceptibility\cite{Smirnov} measurements,
it should be directly confirmed by an x-ray diffraction measurement.

We next discuss complicated behavior experimentally observed near \Tc .
Some experiments suggest the transition is second-order.
X-ray diffraction measurements observed the superlattice
reflection intensity continuously vanish at \Tc\ as
$(T_{\rm C}-T)^{\beta}$ ($\beta\sim$ 0.16).\cite{Ravy,Nakao99}
It was reported that $\nu_c$ for the $^{23}$Na NMR
obeys the power law with $\beta\sim 0.19$.\cite{Fagot}
On the contrary, 
K\"{o}ppen {\it et al}.\ have found two adjacent phase transitions
by a thermal-expansion measurement:
a first-order transition at $T_h\sim 33.0$ K and the other at $T_l\sim 32.7$
K.\cite{Koeppen}
In the $^{51}$V NMR spectrum, the resonance lines in
the uniform and charge-ordered phases were observed to coexist 
in the temperature range between 33.4 and 33.8 K with a similar width 
to the separation between $T_h$ and $T_l$.\cite{Ohama99}
Recent measurements of dielectric and magnetic properties under high pressure
clearly show two separated transitions under pressure.\cite{Sekine}

To investigate these characteristics of the transition,
we measured the temperature variation of
the NMR spectrum for $H\parallel a$ near \Tc\ as 
shown in Fig.~\ref{fig:specA}. 
The experimental error of the absolute value of the temperature is 
less than $0.2$ K.
We found that the lines in the uniform and charge-ordered
phases coexist in a narrow temperature range between 33.4 and 33.8 K
without any hysteresis.
This temperature range agrees with the $^{51}$V NMR measurements 
with a different single-crystalline sample.\cite{Ohama99}
\begin{figure}[tb]
  \begin{center}
\epsfxsize=82mm \epsfbox{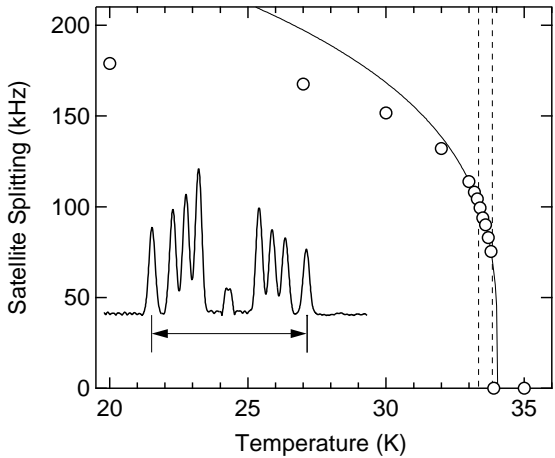}
  \end{center}
\caption{Temperature dependence of satellite splitting for $H\parallel a$.
Solid line is a fitting to the power law $\propto (T_{\rm C}-T)^{\beta}$,
$\beta\sim$ 0.29, and $T_{\rm C}\sim$ 34.04 K.
Dashed lines show the boundaries between which the resonance lines of
the uniform and charge-ordered phases coexist.}
\label{fig:split}
\end{figure}

In Fig.~\ref{fig:split}, we show the temperature dependence of
the satellite splitting for $H\parallel a$.
We can fit this temperature dependence to the power law with $\beta\sim$ 0.29
and $T_{\rm C}\sim$ 34.04 K.
However, the deduced \Tc\ is too high,
 since the low-temperature signal disappears between 33.8 and 33.9 K.
This indicates that the analysis for the ordinary second-order transition
is inadequate.
We therefore conclude that the transition at \Tc\ is first-order.
The present result does not rule out the other adjacent transition,
but its existence is unclear.

In summary,
we have measured the $^{23}$Na NMR spectrum in \navo\ with
a single-crystalline sample.
We have observed eight Na sites in the charge-ordered phase in
disagreement with the low-temperature structure of space group \SG{Fmm2}
determined by the x-ray diffraction measurements.
We discussed possible space groups and charge ordering patterns,
and have concluded that zigzag patterns are the most probable.
We also observed the resonance lines in the uniform and charge-ordered phase
coexist near \Tc . Furthermore,
the temperature dependence near \Tc\ of the satellite line splitting
is inconsistent with that of a second-order transition.
We thus conclude that the transition is first-order.

\section*{Acknowledgment}
We would like to thank H. Nakao, Y. Ohta, Y. Itoh and T. Yamauchi 
for helpful discussions.

\end{document}